\documentclass[conference]{IEEEtran}
\IEEEoverridecommandlockouts
\usepackage{cite}
\usepackage{amsfonts}

\usepackage{textcomp}
\usepackage{xcolor}
\def\BibTeX{{\rm B\kern-.05em{\sc i\kern-.025em b}\kern-.08em
    T\kern-.1667em\lower.7ex\hbox{E}\kern-.125emX}}
    
\usepackage[final]{graphicx}
\graphicspath{ {./} }
\usepackage[reqno]{amsmath}
\usepackage{amssymb}
\usepackage{amsthm}
\usepackage{subfig}
\usepackage{epstopdf}
\usepackage{algorithm}
\usepackage{algorithmicx}
\usepackage{algpseudocode}
\usepackage{setspace}
\usepackage[T1]{fontenc}
\usepackage{color,soul}
\usepackage{multirow}
\usepackage{array}
\usepackage{cite}
\usepackage{verbatim}
\usepackage{enumitem}
\usepackage{url}
\usepackage{csvsimple}
\usepackage{float}

\usepackage{float}
\usepackage{graphicx}
\usepackage{caption}
\usepackage[capitalize]{cleveref}
\usepackage{environ}

\usepackage{array}
\newcommand{\PreserveBackslash}[1]{\let\temp=\\#1\let\\=\temp}
\newcolumntype{C}[1]{>{\PreserveBackslash\centering}p{#1}}
\newcolumntype{R}[1]{>{\PreserveBackslash\raggedleft}p{#1}}
\newcolumntype{L}[1]{>{\PreserveBackslash\raggedright}p{#1}}

\begin{document}

\title{Improving Transferability of Network Intrusion Detection in a Federated Learning Setup}

\author{\IEEEauthorblockN{Shreya Ghosh\IEEEauthorrefmark{1}, Abu Shafin Mohammad Mahdee Jameel\IEEEauthorrefmark{1}, Aly El Gamal\IEEEauthorrefmark{1} }
\IEEEauthorblockA{\IEEEauthorrefmark{1} School of Electrical and Computer Engineering, Purdue University, USA}
\IEEEauthorblockA{Email: {\{ghosh64, amahdeej, elgamala\}@purdue.edu}} }

\maketitle

\begin{abstract}
Network Intrusion Detection Systems (IDS) aim to detect the presence of an intruder by analyzing network packets arriving at an internet connected device. Data-driven deep learning systems, popular due to their superior performance compared to traditional IDS, depend on availability of high quality training data for diverse intrusion classes. A way to overcome this limitation is through transferable learning, where training for one intrusion class can lead to detection of unseen intrusion classes after deployment. In this paper, we provide a detailed study on the transferability of intrusion detection. We investigate practical federated learning configurations to enhance the transferability of intrusion detection. We propose two techniques to significantly improve the transferability of a federated intrusion detection system. The code for this work can be found at https://github.com/ghosh64/transferability.


\end{abstract}
\IEEEpeerreviewmaketitle
\section{Introduction}

A network intrusion detection system, aimed at timely detection of the presence of an intruder, is an essential component for a wide range of network-connected devices. Classic approaches to network intrusion detection depended on techniques like Naive-Bayes classifiers, Random Forest classifiers, and Support Vector Machines \cite{koc2012network}.  

Multiple studies have shown that deep learning based methods provide superior performance compared to traditional approaches \cite{alkasassbeh2016detecting, vinayakumar2019deep}. Recent developments in the IDS domain center around deep learning algorithms \cite{barnard2022robust}, and these approaches are taken as benchmarks during dataset development\cite{koroniotis2019towards}.

Current deep learning models are usually trained to detect a set of known attacks or learn from attacks performed in the past. Although these models are highly accurate in detecting known attacks, they are not always robust against novel attacks. It is difficult to provide guarantees about the kind of attacks that a computer network can expect to see and is susceptible to, especially with the rise in the number of zero-day attacks. Anomaly detection based studies often treat all unknown network intrusions as a single class and as such do not provide a good analysis of model transferability.

\begin{figure*}[h]
    \centering
    \includegraphics[width=1.0\textwidth]{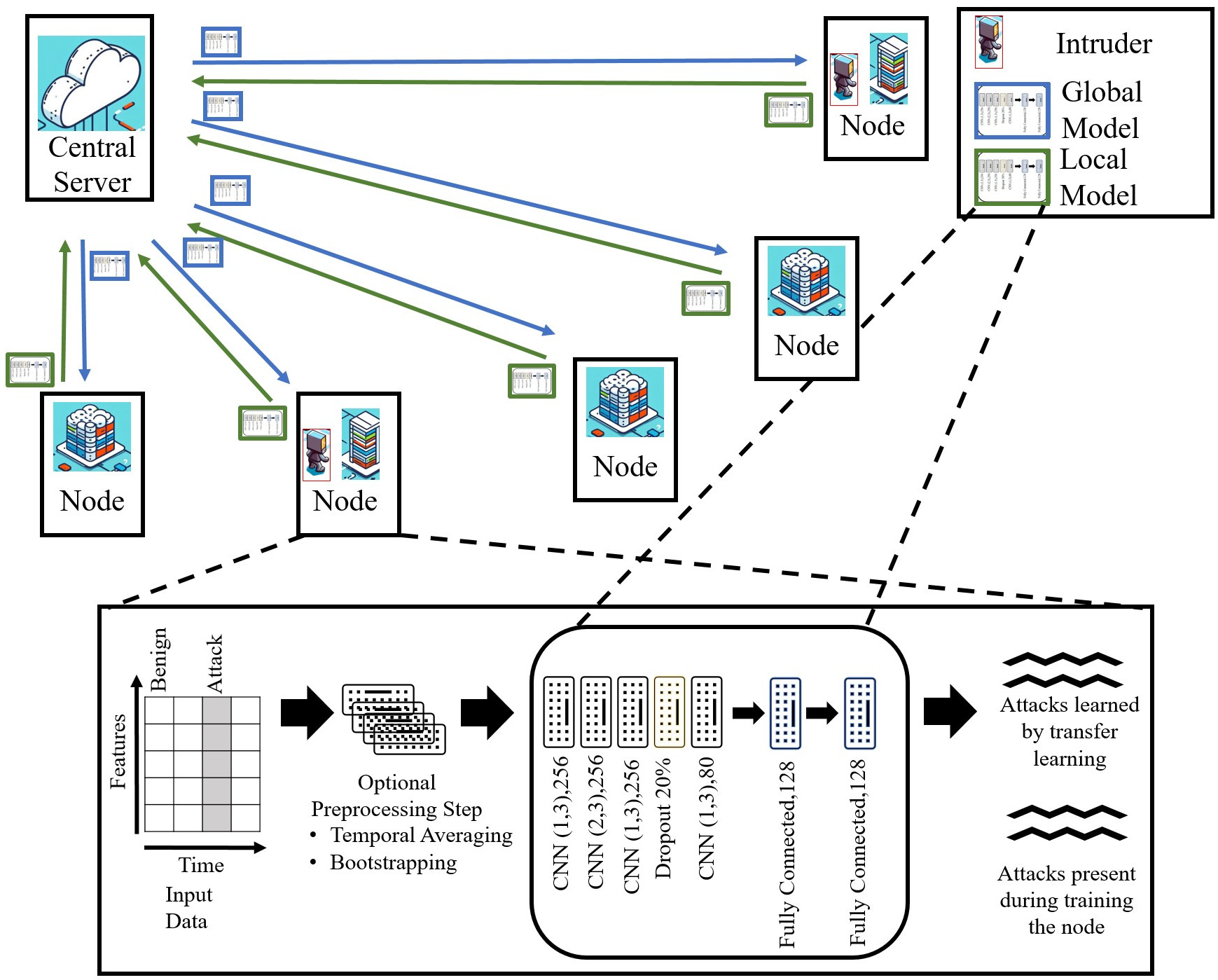}
    \caption{Architecture of proposed intrusion detection model. \vspace{-15pt}}
    \label{fig:sysarch}
\end{figure*}

Model learning, if transferable, would enable these models to be able to detect a wider range of attacks than just the attacks they have been trained to detect. This has resulted in increased focus on transferability studies \cite{verkerken2022towards, catillo2022transferability}. Without transferability studies, it is hard to predict the true range of attacks for which the employed machine learning algorithm is actually robust. Transferability study of a model is concerned with training a deep learning model using data from one source and deploying it to detect data from another source. In \cite{verkerken2022towards}, the authors train their models on one dataset and test it on a separate dataset, with focus on one attack class. The study in \cite{catillo2022transferability} improves on this by focusing on transferability between attack classes, i.e. when the model is trained on one attack class and tested on another. However, the authors pre-select training and testing attack classes. 

Another important recent development regarding data-driven IDS systems has been the shift towards distributed learning. Federated learning approaches to intrusion detection in Internet of Things (IoT) have been shown to outperform non-federated centralized learning by utilizing local knowledge, while preserving data privacy \cite{rahman2020internet}. In a federated setup, the ability to transfer learning from one node to another is of vital importance. There have been significant research interest towards transferability analysis of federated intrusion detection systems \cite{xue2022deep, ruzafa2021intrusion}.

In this paper, we develop a federated deep learning based intrusion detection system, with the explicit aim of enhancing transferablity of intrusion classes between nodes. We propose two pre-processing techniques, bootstrapping and temporal averaging, aimed at improving transferability in a federated learning environment. Our proposed algorithm, which integrates both these approaches, is named \textbf{T}emporally \textbf{A}veraged \textbf{B}ootstrapped \textbf{F}ederated \textbf{I}ntrusion \textbf{D}etection \textbf{S}ystem (TabFIDS). TabFIDS provides a significant improvement in the performance of the federated IDS by considerably increasing the number of observed transferability relationships, as well as boosting the accuracy of transferable intrusion detections. We provide a detailed analysis of the transferability relationships found between different attack classes which demonstrates the superiority of our proposed method.

This paper is organized as follows: In Section \ref{systemarch}, we explain our system architecture, and the federated learning environment. Next, we present the deep learning model developed for intrusion detection. We then explain bootstrapping and temporal averaging, the two pre-processing steps to improve the transferability in a federated system. Finally, in Section \ref{results}, we explain our experimental setup and show that our proposed algorithm significantly improves the transferability performance in a federated environment.

\section{System Architecture}\label{systemarch}

In Fig. \ref{fig:sysarch}, we present a representational scenario illustrating the system environment. In this section, we  discuss different components of this environment.

\subsection{Nodes}
We consider a network of connected devices, treating each device as a separate node. Each node communicates with many other nodes through the network. Malicious actors within this network are attempting to breach the nodes through diverse attack patterns. The nodes are equipped with a deep learning based intrusion detection system that is able to identify attack packets in the network traffic. 

\subsection{Intrusion Detection System}
Before deployment, the IDS on each node has been trained using pre-labeled benign and attack data packets. This training is achieved with the help of a powerful central server. There are two possible roles for this central server:

\subsubsection{Centralized Learning}
In a centralized learning setup, each node sends all its training data to the central server. The central server trains a single deep learning model (same as the local model in Fig. \ref{fig:sysarch}), and sends it back to the nodes. Each node deploys this model without any modification. The nodes do not perform any training.

\subsubsection{Federated Learning}
In a setup employing federated learning, the nodes have individual training capabilities. The nodes use localized training data to train its local IDS model. The local models are then sent to the central server. No training data leaves the node, and the central server gets access to only the model architecture and weights. 

The central server employs a FedAvg aggregation algorithm \cite{fedavg} to generate parameter weights for the global model. This global model is in turn sent back to all the nodes. The nodes then further train this global model using local training data. This process is repeated for multiple communication rounds. The local and global models have the same design, detailed in Fig. \ref{fig:sysarch}, but differ in their parameter weights. 

The operation to calculate global model weights from local model weights can be represented by the following equation:

\begin{equation}
    \forall k, w_{t+1}^k \leftarrow w_t - \eta g_k; w_{t+1} \leftarrow \sum_{k=1}^{K} \frac{n_k}{n} w_{t+1}^k 
\end{equation}
where $w_t$ are the model weights after communication round $t$, $n_k$ is the number of local samples in the training data, assuming we have $K$ total nodes.

\subsection{Deep Learning Model}

We designed the proposed deep learning model through an empirical study maximizing attack data classification accuracy. This architecture is able to perform deep feature extraction using flow based features not only for adequately sampled attacks, but also for severely under represented attacks. In Fig. \ref{fig:sysarch}, we present the architecture of this model. The input is a 1D feature vector containing features extracted from a data packet. This input is then passed through four convolution layers, one dropout layer and two fully connected layers. The output is one-hot encoded, detecting benign or attack data samples. We use the same model in our analysis of centralized and federated environments. The transferability of this model can be measured by its ability to detect test attacks that were not present during the training phase. Our goal here is to develop a model that not only performs well in localized testing (testing with attack classes it has seen during training), but also excels in transferability (when tested with attack classes it has never encountered during the training phase).

\subsection{Bootstrapping}

One feature of network intrusion is that an overwhelming majority of data packets encountered by a node are benign data packets. In comparison, the intrusion attacks generate very few data samples. This makes it challenging to design an efficient deep learning based IDS, as deep learning performance is heavily dependent on the availability of ample training samples. One popular technique to create a more balanced dataset is to bootstrap the training data. By repeatedly re-sampling the original minority class training samples (in this case, the attack samples), a balanced training dataset is created. 

\subsection{Temporal Averaging}

Temporal averaging, when employed as a pre-processing step, has been shown to boost the transferability of an IDS system \cite{ghosh2023an}. In this case, each input sample to the model is the temporal average of $r$ other samples. Mathematically this can be given as:

\begin{equation}
    y_{t}=\frac{\sum_{i=0}^{r-1} x(t-i)}{r}, r=\text{window size} 
\end{equation}
where $x_{t}$ represents the input data sample at time $t$ and $y_{t}$ represents the corresponding temporally averaged data sample. $y_{t}$ is then fed to the deep learning IDS model. Temporal averaging further improves privacy of the data, as the deep learning model is trained with the pre-processed data and it never sees the original data.

\begin{figure*}[t]

    \centering
    \includegraphics[width=0.8\textwidth]{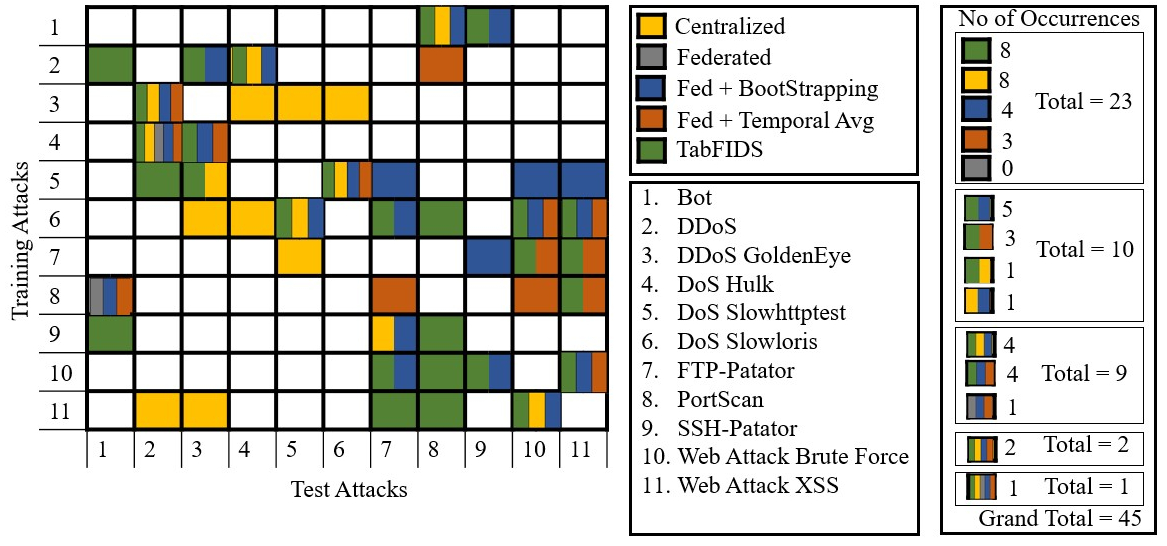}
    \caption{A summary of transferable train-test attack pairs. Colors denote which approach features a specific transferability pair. In case of a pair being present in multiple approaches, striped coloring is utilized.}
    
    \label{fig:transferability_overview}
    
\end{figure*}

\section{Experimental Results} \label{results}

\subsection{Dataset}

Research in this domain has led to the availability of multiple standardized network traffic datasets, generated using traffic monitoring software to log and monitor network traffic packets. These datasets have a high dimensional feature space, making it non-trivial for machine learning models to learn meaningful information. 

The dataset used to train our model is generated by the Canadian Institute of Cybersecurity: CIC-IDS 2017 \cite{sharafaldin2018toward}. It consists of 78 features and 14 types of attacks. The classes in this dataset are imbalanced - while 80\% of the data is benign, only 20\% of the data is from attack packets. We ignore three attack classes that suffer from extremely low data availability, together containing only .00232\% of total data. Even then, the lowest remaining attack class has .023\% of total data, making it extremely challenging for a data-driven IDS. We use 80\% of this data for training, 10\% for validation and 10\% for testing.

\subsection{Experimental Setup}

Firstly, in a centralized environment, the  model is trained on all available benign data in the training set and one class of attack data. Then, the transferability is evaluated by testing the model against test data packets from attack classes it has not seen during the training phase. We perform this experiment for all 11 attack classes.

The main aim of this paper is to analyze and improve transferability in a federated setup. In this environment, we divide the training dataset to 11 separate nodes. In order to better understand transferability of individual attacks from one node to another, during training, we expose each node to a single attack type only. There is also no overlap between the benign data present in different nodes during training. Each node only sees a $\frac{1}{11}$ fraction of the training data, effectively making local training data set availability much lower compared to a centralized system. 

When employing bootstrapping, we sample attack data packets from the same class with replacement until the dataset contains 50\% benign data and 50\% attack data. We use a window size of 3 for temporal averaging. We employ the Adam\cite{adam} optimizer with a learning rate of 0.001 and the loss function is CrossEntropy Loss. For the federated model, we run 20 rounds of federated model aggregation.

\subsection{Evaluation Metric} \label{subsec:evaluation}
As mentioned before, intrusion detection systems encounter very few attack data packets and a lot of benign data packets. This poses an unique problem regarding the evaluation data metric. As the IDS is generally assumed to be efficient in classifying benign data packets, its classification accuracy might be high even if it fails at its main task of identifying intrusion data packets. For example, in a dataset containing 90\% benign data, a classifier may achieve 90\% accuracy even if it fails to detect even a single attack data packet. 

To address this problem, a lot of the relevant investigations involving an imbalanced dataset report the precision and recall values. Assume $tp$, $tn$, $fp$ and $fn$ represent accurately detected attack data points, accurately detected benign data points, wrongly detected attack data points, and wrongly detected benign data points, respectively. Then the overall accuracy of this system would be $(tp+tn)/(tp+tn+fp+fn)$, the precision would be $(tp)/(tp+fp)$ and recall would be $(tp)/(tp+fn)$. The recall value provides us with the ratio of attack data samples that are accurately classified. However, it does not contain any information about whether the system can accurately identify the benign data samples. 

To address this issue, we propose to use the attack accuracy, defined as:

\begin{equation}
    Accr=\frac{\frac{tn}{tn+fp}+\frac{tp}{tp+fn}}{2}
\end{equation}
which gives equal weight to correct detection of benign and attack data packets. 

\begin{table}[t]
        \centering
        \captionsetup{justification=centering}
	\caption{Number of Occurrences for an Attack Class in Transferable (Train Attack, Test Attack) Pairs.}
	\label{table:train-test-stats}
    \tabcolsep=0.12cm
    \begin{tabular}{l|c|c|c|c|c|c|c|c|c|c|c} 
    \hline
     Attack Number& \textbf{1}& \textbf{2}& \textbf{3}&\textbf{4} & \textbf{5}& \textbf{6}& \textbf{7}& \textbf{8}&\textbf{9} & \textbf{10}&\textbf{11}\\
     \hline
     Present as Train Attack& 2& 4& 4&2& 6& 7& 4& 4& 3& 4&5\\ 
    \hline
    Present as Test Attack& 3& 4& 5&3& 3& 2& 6& 6& 3& 5&5\\
    \hline
    \end{tabular}
    \vspace{-15pt}
    
\end{table}

\begin{figure*}[t]
    \centering
    \includegraphics[width=0.8\textwidth]{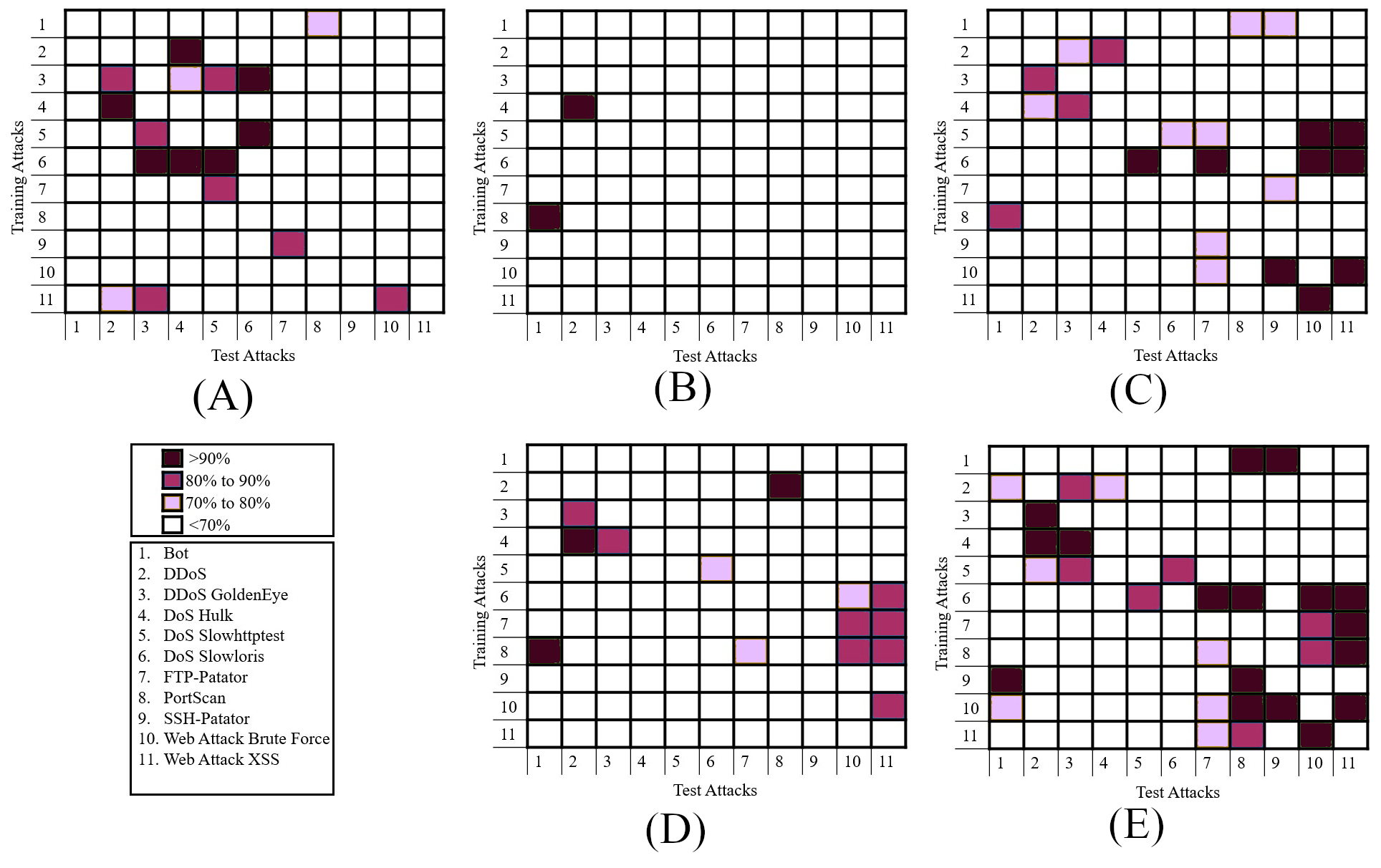}
    \caption{Details of transferable train-test attack pairs across different approaches (A) Centralized, (B) Federated, (C) Federated with Bootstrapping, (D) Federated with Temporal Averaging, and (E) TabFIDS.}
    
    \label{fig:transferability_percentage}

\end{figure*}

\subsection{Transferability Relationships}

In Fig. \ref{fig:transferability_overview}, we present a summary of our findings. Here, we investigate transferability relationships for the centralized and federated approaches. We further illustrate the transferability pairs that occur when the federated approach is augmented with bootstrapping or temporal averaging. Finally, we also exhibit results for an approach where both bootstrapping and temporal averaging are incorporated in the pre-processing step. In the figure, we use color codes to identify approaches that generate transferable pairs between a pair of training and testing attack classes. If a pair is transferable through multiple approaches, it is colored with stripes identifying each approach. 

For the purpose of this study, we define transferability as an IDS trained on one attack class being able to achieve an attack accuracy greater than 70\% when tested on another attack class. The training attack class, and the test attack class then belong to a transferability pair, and each transferability pair is denoted by (training attack number, test attack number). The type of attack represented by each attack number is included in Fig. \ref{fig:transferability_overview}. We summarize some of the findings from Fig. \ref{fig:transferability_overview} below:

\begin{itemize}
    \item The existence of a colored block in a row is evidence of a node trained with that attack class being able to detect other unseen attack classes. Similarly, the existence of a colored block in a column is evidence of that attack class being in a detected (train, test) transferability pair as a test attack. The diagonals of this plot represent cases when training and testing is done for the same attack class. All the considered approaches perform well in this case, and we have chosen to ignore these cells in the figure, in order to highlight the transferability pairs.

    We can see that, each row of the plot has at least two colored blocks, and each column also has at least two colored blocks. This means that all attack classes are present in at least two transferability pairs. In Table. \ref{table:train-test-stats}, we present a detailed analysis of the number of occurrences for an attack class in transferability pairs. Attack 6, DoS Slowloris shows the highest amount of transferability as a training class. Attacks 7 and 8, FTP-Patator and PortScan are the attacks most commonly being detected in nodes trained on other attack classes.

    \item In the right side of the Fig. \ref{fig:transferability_overview}, we summarize the distribution of the transferability pairs. Using all possible approaches, there is a grand total of 45 transferable pairs, out of 110 possible pairs. Of them, 23 show transferability under one approach only. On the other hand, for only one transferability pair (DoS Hulk, DDoS), transferability can be uncovered by all five approaches.

    \begin{table}[t]
        \centering
        \captionsetup{justification=centering}
	\caption{Number of Transferable Pairs for Different Approaches and their Attack Accuracy.}
	\label{table:transferable-pairs}
    \tabcolsep=0.05cm
    \begin{tabular}{l|c|c|c|c} 
    \hline
     & Total& >90\%& 80\%-90\%&70\%-80\%\\
     \hline
     Central& 17& 7& 7&3\\ 
    \hline
    Federated& 2& 2& 0&0\\ 
    \hline
    Fed+BStrap&     22&     9&      4&9\\ 
    \hline
    Fed+TempAv&    14&    3&     3&8\\
    \hline
    TabFIDS& \textbf{31}& 17& 7&7\\
    \hline
    \end{tabular}
    \vspace{-15pt}
    
\end{table}
    
    \item Then in Fig. \ref{fig:transferability_percentage}, we present an analysis of the five approaches in terms of transferable pairs, and then break them down by transferability performance. This figure is useful in identifying transferability pairs that show a very high (above 90\%), high (80\% to 90\%) or moderate (70\% to 80\%) transferability in terms of attack accuracy. The findings from Fig. \ref{fig:transferability_percentage} are summarized in Table \ref{table:transferable-pairs}. 
    
    Firstly, we note that the centralized model performs reasonably well, detecting 17 transferable pairs. Then we shift our attention to the federated approaches. One advantage of federated learning is that there is exchange of information between different nodes with different training attacks during aggregation, and therefore, the final local model can be expected to perform well in terms of transferability. However, when we deploy the same deep learning model in a federated learning environment, there is a dramatic reduction in transferablity performance, with only 2 transferability pairs being detected. One reason for this can be that the individual training sets of each node are smaller compared to the centralized training set, and that might amplify the effects of the imbalanced dataset.

    \item Next, we augment federated learning with bootstrapping, which should aid in overcoming the effects of the unbalanced dataset. From the results in Table \ref{table:transferable-pairs}, we can see that this results in significantly improved performance, with 22 transferable pairs.

    \item We also conduct separate tests on the impact of temporal averaging on transferability. This approach leads to a significant enhancement compared to the federated setup, with 14 transferability pairs compared to 2. This is a remarkable result, as it shows that even in the presence of a serious imbalance in a dataset, temporal averaging can improve the performance of a deep learning model without resorting to data augmentation. This can be due to the fact that temporal averaging averages multiple input data packets, providing the IDS with more information about the network environment. This can be very useful in cases where training is resource-constrained, as data augmentation considerably increases the training time.

    \item Finally, we combine both bootstrapping and temporal averaging. This proposed approach, named \textbf{T}emporally \textbf{A}veraged \textbf{B}ootstrapped \textbf{F}ederated \textbf{I}ntrusion \textbf{D}etection \textbf{S}ystem (TabFIDS), provides by far the best results in terms of transferability, achieving transferability for 31 pairs. From Fig. \ref{fig:transferability_percentage} and Table \ref{table:transferable-pairs}, we further note that for pairs where transferability is achieved, TabFIDS has the most number of pairs with an attack accuracy greater than $90\%$. While TabFIDS has 17 such pairs, the next best approach has only 9 such pairs. 
    We also observed that TabFIDS performs exceptionally well in localized intrusion detection too, achieving an average accuracy of $95.36\%$ across nodes, when a node is tested using the same attack class encountered during training.
\end{itemize}

Interestingly, in the process of combining temporal averaging and bootstrapping for TabFIDS, we lose the ability to detect some transferability pairs that could be detected by standalone bootstrapping or temporal averaging based approaches. From Fig. \ref{fig:transferability_overview}, we can see that there are 7 such pairs, 4 from bootstrapping and 3 from temporal averaging. This indicates that there might be further room for improving TabFIDS by designing better integration methods between temporal averaging and bootstrapping. Similarly, there are 8 transferability pairs that are uniquely detected by only the centralized model. There is scope to investigate the reason for this, which might result in further improvement of TabFIDS.

\begin{figure}[t]
    \centering
    \includegraphics[width=0.45\textwidth]{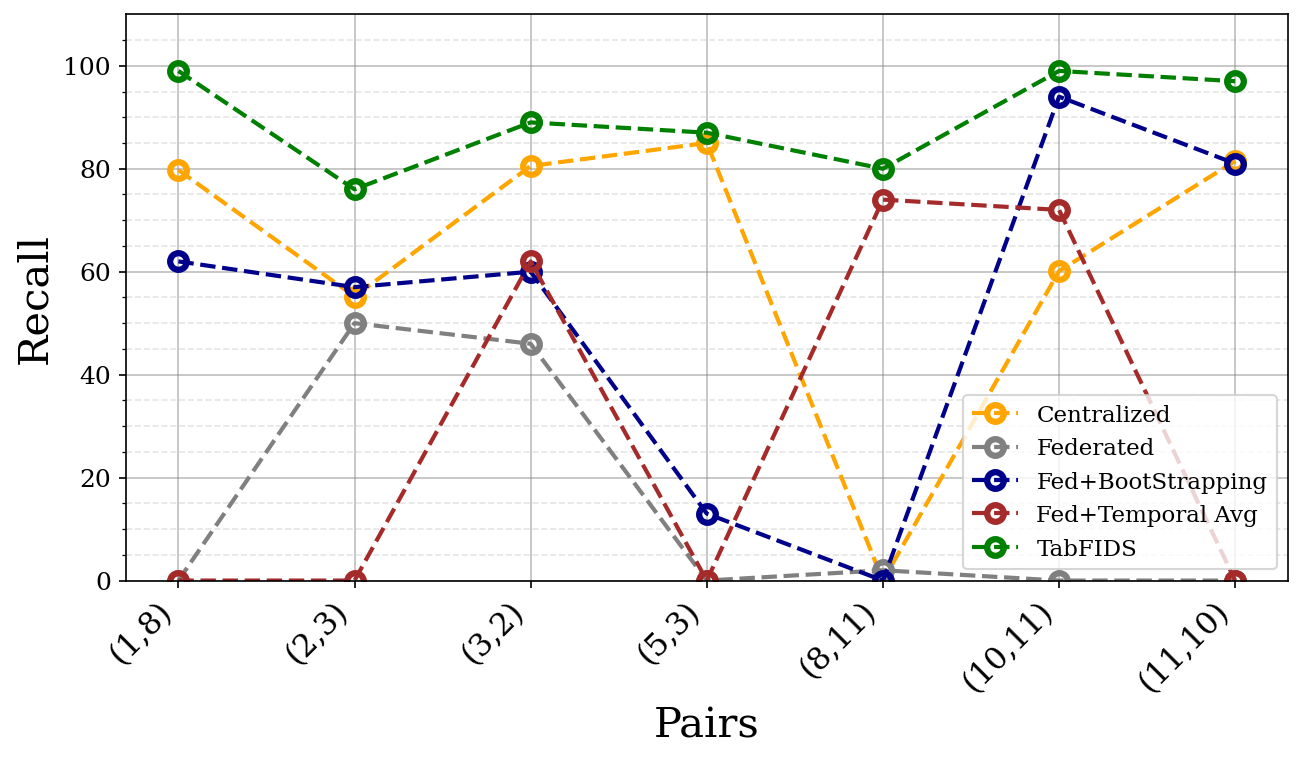}
    \caption{Selected cases where TabFIDS improves transferability.}
    \label{fig:compexisting}
\end{figure}

\begin{figure}[t]
    \centering
    \includegraphics[width=0.45\textwidth]{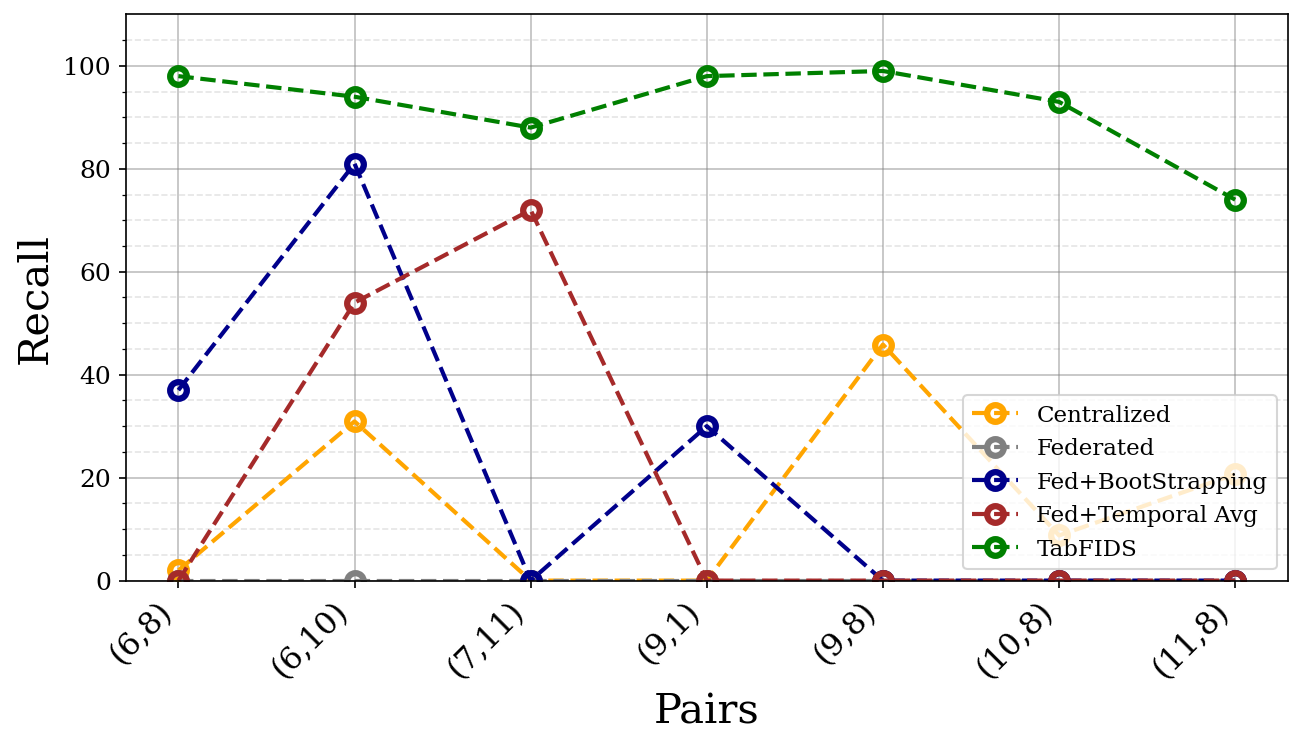}
    \caption{Selected cases where TabFIDS uncovers transferability. }
    \label{fig:compnew}
\end{figure}

In Fig. \ref{fig:compexisting}, we take a closer look at the ability of TabFIDS to improve the accuracy of transferable intrusion detection using the recall values. Specially in the case of (5,3), we can see that TabFIDS uncovers transferability even when all other federated methods fail. In other cases, TabFIDS improves the transferability performance. Then, in Fig. \ref{fig:compnew}, we focus specifically on transferability pairs where other approaches struggle with detection of the attack class.  Specially in the cases of PortScan as a test attack ((9,8), (10,8), and (11,8)), TabFIDS uncovers transferability which none of the other methods can do. Employing recall as a metric helps highlight the limitations of certain approaches that struggle to accurately classify even a single attack data sample in specific transferability pairs.

\section{Conclusion}

In this work, we develop a federated deep learning based intrusion detection system with enhanced transferability. We evaluate bootstrapping and temporal averaging, two pre-processing techniques, whose combination results in a significant increase in transferability performance. While originally designed for intrusion detection systems, the TabFIDS model developed here can be of value in other deep learning domains where transferability is a desired attribute. Opportunities remain for enhancing TabFIDS through intelligent integration of bootstrapping and temporal averaging.

\ifCLASSOPTIONcaptionsoff
  \newpage
\fi
\bibliographystyle{IEEEtran} 

\bibliography{nattack2021}

\end{document}